\documentclass{article}

\usepackage{amsfonts}
\usepackage{amssymb}
\usepackage{pseudocode}
\usepackage{listings}
\usepackage{hyperref}
\usepackage{amsthm}
\usepackage{latexsym}
\usepackage{xspace}
\usepackage{graphicx}
\usepackage{amsmath}
\usepackage[latin1]{inputenc}
\usepackage{epsfig}
\usepackage{mathrsfs}

\newtheorem{ex}{Example}[section]

\newcommand{\bp}{\boldsymbol{p}}

\newcommand{\bs}{\boldsymbol{s}}

\newcommand{\bx}{\boldsymbol{x}}

\newcommand{\bm}{\boldsymbol{m}}
\newcommand{\by}{\boldsymbol{y}}

\newcommand{\bg}{\boldsymbol{g}}

\newcommand{\be}{\boldsymbol{e}}
\newcommand{\bzero}{\boldsymbol{0}}

\begin{document}

\title{Locally epistatic genomic relationship matrices for genomic association, prediction and selection}
\author{Deniz Akdemir \\ Department of Plant Breeding \& Genetics\\ 
  Cornell University\\ Ithaca, NY}

\maketitle

\begin{abstract} 
In plant and animal breeding studies a distinction is made between the commercial value (additive + epistatic genetic effects) and the breeding value (additive genetic effects) of an individual since it is expected that some of the epistatic genetic effects will be lost due to recombination. In this paper, we argue that the breeder can take advantage of some of the epistatic marker effects in regions of low recombination. The models introduced here aim to estimate local epistatic line heritability by using the genetic map information and combine the local additive and epistatic effects. To this end, we have used semi-parametric mixed models with multiple local genomic relationship matrices with hierarchical testing designs and lasso post-processing for sparsity in the final model and speed. Our models produce good predictive performance along with genetic association information.   
\end{abstract}


{Keywords \& Phrases: Genomic selection, Genome wide association, Plant / animal breeding, Mixed model, Multiple kernel learning, Heritability}

\section{Introduction}

Selection in animal or plant breeding is usually based on estimates of genetic breeding values (GEBV) obtained with semi-parametric mixed models (SPMM). In these mixed models genetic information in the form of a pedigree or markers are used to construct an additive kernel matrix that describes the similarity of line specific additive genetic effects. These models have been successfully used for predicting the breeding values in plants and animals. The studies show that using similarities calculated from sufficient genome wide marker information almost always lead to better prediction models for the breeding values compared to the pedigree based models. In both simulation studies and in empirical studies of dairy cattle, mice and in bi-parental populations of maize, barley and \emph{Arabidopsis} marker based SPMM GEBVs have been quite accurate.

A SPMM for the $n\times 1$ response vector $\by$ is expressed as 
\begin{equation}\label{eq:spmm} \by=X\beta+Z\bg+\be \end{equation} where $X$ is the $n\times p$ design matrix for the fixed effects, $\beta$ is a $p\times 1$ vector of fixed effects coefficients, $Z$ is the $n\times q$ design matrix for the random effects; the random effects $(\bg',\be')'$ are assumed to follow a multivariate normal distribution with mean $\bzero$ and covariance \[ \left( \begin{array}{cc}
\sigma^2_g K  & \bzero  \\
\bzero & \sigma^2_e I_n \end{array} \right)\] where $K$ is  a $q\times q$ kernel matrix.

The kernel of the marker based SPMM's and reproducing kernel Hilbert spaces (RKHS) regression models have been stressed recently (\cite{gianola2008reproducing}). In fact, the connection have been recognized long time ago by \cite{kimeldorf1970correspondence}, \cite{harville1983discussion}, \cite{robinson1991blup} and \cite{speed1991blup}. RKHS regression models use an implicit or explicit mapping of the input data into a high dimensional feature space defined by a kernel function. This is often referred to as the ''kernel trick'' (\cite{scholkopflearning}).  It is possible to say that RKHS regression extends SPMM's by allowing a wide variety of kernel matrices, not necessarily additive, calculated using a variety of kernel functions. The common choices for kernel functions are the linear kernel function, polynomial kernel function, Gaussian kernel function though many other options are available.

A kernel function, $k(.,.)$ maps a pair of input points $\bx$ and $\bx'$ into real numbers. A kernel function is by definition symmetric ($k(\bx,\bx')=k(\bx',\bx)$) and non-negative. Given the inputs for the $n$ individuals we can compute a kernel matrix $K$ whose entries are $K_{ij}=k(\bx_i,\bx_j).$ The linear kernel function is given by $k(\bx; \by) = \bx'\by.$ The polynomial kernel function is given by $k(\bx; \by) =(\bx'\by+ c)^d$ for $c$ and  $d$ $\in$ $R.$  Finally, the Gaussian kernel function is given by $k(\bx; \by) = exp(-(\bx'-\by)'(\bx'-\by)/h)$ where $h>0.$ Taylor expansions of these kernel functions reveal that each of these kernels correspond to a different feature map.

For the marker based SPMM's, a genetic kernel matrix calculated using a linear kernel matrix incorporates only additive effects of markers. A genetic kernel matrix based on the polynomial kernel of order $k$ incorporates all of the one to $k$ order monomials of markers in an additive fashion. The Gaussian kernel function allows us to implicitly incorporate the additive and complex epistatic effects of the markers. 

Simulation studies and results from empirical experiments show that the prediction accuracies of models with Gaussian or polynomial kernel are usually better than the models with linear kernel. However, it is not possible to know how much of the increase in accuracy can be transfered into better generations because some of the predicted epistatic effects that will be lost by recombination. This issue touches the difference between the commercial value of a line which is defined as the overall genetic effect (additive+epistatic) and the breeding value which is the potential for being a good parent (additive) and it can be argued that linear kernel model estimates the breeding value where as the Gaussian kernel estimates the commercial value. In this article, we argue that the breeder can take advantage of some of the epistatic marker effects in regions of low recombination. The models introduced here aim to estimate local epistatic line heritability by using the genetic map information and combine the local main and epistatic effects. Since the epistatic effects that are incorporated are only local there is little chance that these effects will disappear with recombination. Heritability is defined as the percentage of total variation that can be explained by the genotypic component. One can similarly argue that SPMM's with linear kernels produce estimates of narrow sense line heritability, and the SPMM's with Gaussian Kernel produces estimates of broad sense line heritability. We expect that the estimates of local heritability developed in this paper to be between narrow and broad sense heritability. 

An issue with the over the shelf kernel functions like linear or Gaussian kernels is that same kernel matrix is used no matter what trait is considered and that all markers are assigned equal weighs in the analysis. When the relation matrix is not task-specific, it is often one that forces the solution to be overly smooth.

We propose several approaches for local kernel matrix calculation. Our final models are SPMM's with semi-supervised kernel matrix that is obtained as a function of many local kernels. They differ mainly in the way the local kernel matrices and their weights are calculated. One major aim of this article is to measure and incorporate additive and local epistatic genetic contributions since we believe that the local epistatic effects are relevant to the breeder.

The local heritability models in this article can be adjusted so that genetic contribution of the whole genome, the chromosomes, or local regions can be obtained. In the following sections, we will discuss several ways in which this information can be useful to the breeder and we will illustrate a breeding scheme where the local instead of genome wide effects are utilized.

\section{Multiple kernel learning with SPMM}

In recent years, several methods have been proposed to combine multiple kernel matrices instead of using a single one. These kernel matrices may correspond to using different notions of similarity or may be using information coming from multiple sources. For example, genomic kernel + pedigree kernel, chromosome model, linear mixed models with linear covariance structure.  A good review and taxanomy of multiple kernel learning algorithms in the machine learning literature can be found in \cite{gonen2011multiple}. 

Multiple kernel learning methods use multiple kernels by combining them into a single one via a combination function. The most commonly used combination function is linear. Given kernels $K_1, K_2,\ldots ,K_p,$ a linear kernel is of the form \[K=\eta_1K_1+\eta_2 K_2+\ldots+\eta_p K_p.\] The kernel weights $\eta_1, \eta_2, \ldots, \eta_p$ are usually assumed to be positive and this corresponds to calculating a kernel in the combined feature spaces of the individual kernels. We will also assume that the weights sum to one.

The components of $K$ are usually input variables from different sources or different kernels calculated from same input variables. The kernel $K$ can also include interaction components like $K_i\odot K_j,$  $K_i\otimes K_j,$ or perhaps $-(K_i-K_j)\odot (K_j-K_i).$ For example, if $K_E$ is the environment kernel matrix and $K_G$ is the genetic kernel matrix, then a component $K_E\odot K_G$ can be used to capture the gene by environment interaction effects.

The mixed models in \cite{burgueno2007modeling} use $A\odot A$ to capture interaction effects. The reasoning comes from (Falconer and Mackay, 1996):

''If one assumes no dominance, all terms will vanish except the terms for additive and additive x additive variances, which will take the form $C_{i'i}=2f_{i'i}\sigma^2_a+(2f_{i'i})^2\sigma^2_{aa}$ 
where $f_{i'i}$ is the COP between individuals $i'$ and $i$, $\sigma^2_a$ is the additive genetic variance, and $\sigma_{aa}$ is the additive x additive genetic variance. Assuming linkage and identity equilibrium, it seems justified to use $(2f_{i'i})^2$, which in matrix notation can be represented by $(A\odot A)=\tilde{A}$ as the coefficient of the additive x additive component'' (where $\odot$ is the element-wise multiplication operator).''

Although some multiple kernel approaches use fixed weights for combining kernels, in most cases the weight parameters need to be learned from the training data. Some principled techniques used to estimate these parameters include likelihood based approaches in the mixed modeling framework like Fisher scoring algorithm or variance least squares approach though these approaches are more suitable to cases where only a few kernels are being used. 

\cite{qiu2009framework} propose two simple heuristics to select kernel weights in regression problems:
\[\eta_m=\frac{r^2_m}{\sum_{h=1}^p{r^2_h}}\] and
\[\eta_m=\frac{\sum_{h=1}^p{M_h}-M_m}{(1-p)\sum_{h=1}^p{M_h}}\]
where $r_m$ is the Pearson correlation coefficient between true response and the predicted response and $M_m$ is the mean square error generated by the regression using the kernel matrix $K_m$ alone. Another approach in \cite{qiu2009framework} uses the kernel alignment: \[\eta_m=\frac{A(K_m, \by\by')}{\sum_{h=1}^p{A(K_h, \by\by')}}\]
where kernel alignment is calculated using \[A(K_1, K_2)=\frac{\left\langle K_1, K_2 \right\rangle_F}{\sqrt{\left\langle K_1, K_1 \right\rangle_F \left\langle K_2, K_2 \right\rangle_F}}\] and \[\left\langle K_1, K_1 \right\rangle_F =\sum_{i,j=1}^{N}(K_1)_{ij} (K_2)_{ij}.\]

\subsection{Multiple kernel SPMM models}
In the context of the SPMM's we propose using weights that are proportional to the estimated variances attributed to the kernels. One possible approach is to use a SPMM with multiple kernels in the form of 
\begin{equation}\label{eq:spmmmk} \by=X\beta+Z_1\bg_1+Z_2\bg_2+\ldots+Z_k\bg_k+\be \end{equation} where $\bg_j\sim N_{q_k}(\bzero, \sigma^2_{g_j}K_j)$ for $j=1,2,\ldots,k.$ Let $\hat{\sigma}^2_{g_j}$ for $j=1,2,\ldots,k$ and $\hat{\sigma}^2_{e}$ be the estimated variance components. Under this model calculate heritabilities as $h^2_m=\hat{\sigma}^2_{g_m}/(\sum_{\ell=1}^k{\hat{\sigma}^2_{g_\ell}}+\hat{\sigma}^2_{e})$ for $m=1,2,\ldots,k.$ 

Another model incorporates the marginal variance contribution for each kernel matrix.  For this we use the following SPMM:
\begin{equation}\label{eq:spmmmk2} \by=X\beta+Z_j\bg_j+Z_{-j}\bg_{-j}+\be \end{equation} where $\bg_j\sim N_{q_k}(\bzero, \sigma^2_{g_j}K_j)$ for $j=1,2,\ldots,k.$ $\bg_{-j}$ is the random effect corresponding to the input components other than the ones in group $j.$  In this case calculate heritabilities as $h^2_m=\hat{\sigma}^2_{g_m}/(\hat{\sigma}^2_{g_m}+\hat{\sigma}^2_{g_{-m}}+\hat{\sigma}^2_{em})$ for $m=1,2,\ldots,k.$ In our illustrations it was always $Z=Z_1=Z_2= \ldots= Z_k,$ however the above models apply to more general cases. 

A simpler approach is to use a separate SPMM for each kernel. Let $\hat{\sigma}^2_{g_m}$ and $\hat{\sigma}^2_{e_m}$ be the estimated variance components from the SPMM model in (\ref{eq:spmm}) with kernel $K=K_m.$ Let $h^2_m=\hat{\sigma}^2_{g_m}/(\hat{\sigma}^2_{g_m}+\hat{\sigma}^2_{e_m}).$  Note that, in this case, the markers corresponding to the random effect $\bg_{-j}$ which mainly accounts for the sample structure can now be incorporated by a fixed effects via their first principal components. 

After heritabilities are obtained, calculate the kernel weights $\eta_1, \eta_2, \ldots, \eta_p$ as \begin{equation}\eta_m=\frac{h^2_m}{\sum_{h=1}^p{h^2_h}}.\end{equation} 

The estimates of parameters for models in (\ref{eq:spmm}), (\ref{eq:spmmmk}) and  (\ref{eq:spmmmk2}) can be by maximizing the likelihood or the restricted (or residual, or reduced) maximum likelihood (REML). There are very fast algorithms devised for estimating the parameters of the single kernel model in  (\ref{eq:spmm}). However, an advantage with the multiple kernel approach in models  (\ref{eq:spmmmk}) and  (\ref{eq:spmmmk2}) is that they can be used for testing nested models through the likelihood ratio test. Estimating the parameters of Model (\ref{eq:spmmmk}) gets very difficult with large number of kernels and with large sample sizes, the single kernel or the marginal kernel models are more suitable in such cases.  
 
\subsection{Hierarchical testing for sparsity and speed}

Although somewhat different in all of the above models, the kernel weights $\eta_1, \eta_2, \ldots, \eta_p$ can be interpreted as the contribution components to the response variable. In the context of Model (1) certain tests are devised for testing whether $\sigma^2_g$ is zero against the one sided alternative. 

Under Model (1), twice the log-likelihood of $\by$ given the parameters $\beta,$ $\sigma^2_e$ and $\lambda=\sigma^2_g/\sigma^2_e$ is, up to a constant,
\begin{equation}L(\beta, \sigma^2_e, \lambda)=-n \log{\sigma^2_e}-\log |V_\lambda|-\frac{(\by-X\beta)'V^{-1}_\lambda(\by-X\beta)}{\sigma^2_e}\end{equation} where $V_\lambda=I_n+\lambda ZKZ'$ and $n$ is the size of the vector $\by.$ 

Twice the residual log-likelihood (\cite{robinson1987program}, \cite{harville1977maximum}) is, up to a constant,

\begin{equation}RL(\beta, \sigma^2_e, \lambda)=-(n-p) \log{\sigma^2_e}-\log |V_\lambda|-\log (X'V_\lambda X)-\frac{(\by-X\hat{\beta}_\lambda)'V^{-1}_\lambda(\by-X\hat{\beta}_\lambda)}{\sigma^2_e}\end{equation} 

From both of these likelihoods, a test statistic for the significance of the variance component 
\begin{eqnarray*}
  H_0 & : & \quad \lambda = 0 \quad  (\sigma^2_g=0) \quad \\
  H_A & : & \quad \lambda > 0 \quad  (\sigma^2_g>0)
\end{eqnarray*}
can be obtained by calculating the likelihood ratio statistic (\cite{cox19741heoretical}). 

Under standard regularity conditions the null distribution of the likelihood ratio test statistic has a $\chi^2$ distribution with degrees of freedom given by the difference in the number of parameters between the null and alternative hypothesis. However, \cite{self1987asymptotic} showed that the asymptotic distribution is a weighted mixture of $\chi^2$ distributions.  For the SPMM in (1) they have recommended using a equally weighted mixture of $\chi^2(0)$ and $\chi^2(1)$ distribution where $\chi^2(0)$ distribution refers to a distribution degenerate at $0.$ In simulation studies \cite{pinheiro2000mixed}, \cite{morrell1998likelihood} found that equal contributions work well with residual log-likelihood where as a $0.65$ to $0.35$ mixture works better for the log-likelihood.  A finite sample null distribution was recommended for the SPMM in (1) in \cite{crainiceanu2003likelihood} and it was shown that the mixture proportions depended on the kernel matrix. 

In many practical cases a thresholding method can be sufficient for the purposes of identifying regions that contribute to phenotype variation. Relevant regions are divided further subregions and the procedure is repeated to a desired detail level. Nevertheless, suitable hierarchical testing procedures have been developed. \cite{blanchard2005hierarchical} proposed and analyzed several hierarchical designs in terms of their cost / power properties.  Multiple testing procedures where coarse to fine hypotheses are tested sequentially  have been proposed to control the family wise error rate or false discovery rate (\cite{reiner2003identifying}, \cite{meinshausen2008hierarchical}).  These procedures can be used along the ''keep rejecting until first acceptance'' scheme to test  hypotheses in an hierarchy. 

Meinshausen's hierarchical testing procedure controls the family wise error  by adjusting the significance levels of single tests in the hierarchy. The procedure starts testing the root node $H_0$ at level $\alpha.$ When a parent hypothesis is rejected one continues with testing all the child nodes of that parent. The significance level to be used at each node $H$ is adjusted by a factor proportional to the number of variables in that node: \[\alpha_H=\alpha\frac{|H|}{|H_0|}\] where $|.|$ denotes the cardinality of a set. This means that larger penalty is incurred at finer levels. The inheritence procedure in \cite{goeman2010inheritance} provides a uniform improvement over the method by Meinshausen. Two hypothetical hierarchical tests are displayed in Figures \ref{fig:minipage1}. and \ref{fig:minipage2}.

\subsection{A multiple kernel model with lasso penalty for sparsity}

Although we can include sparsity in our multiple kernel model by use of hierarchical testing procedures described in the previous section, we can also accomplish this by means of a general additive model with lasso penalty post-processing formulation. 

Each multiple local kernel SPMM model discussed in previous section can be utilized to obtain EBLUPs from the specific regions. Let $\bx$ be the $p$ vector of fixed effects and $\bm=(\bm_1,\bm_2,\ldots,\bm_k)$ be the vector of markers partitioned into k regions. Let $\hat{g}_{j}(\bm)$  denote the EBLUPs of random effect components that correspond to the $k$ local kernels for regions $j=1,2,\ldots, k$ and individual with markers $\bm.$ Consider a final prediction model in the following form:
\begin{equation}f(\bx,\bm; \beta, \alpha)=\beta_0+\sum_{j=1}^{k}\alpha_{j}\hat{g}_{j}(\bm)+\sum_{j=k+1}^{k+p}\beta_j x_j.\label{eq:additivemodel}\end{equation}

Estimate the model coefficients using the following loss function \begin{equation}(\hat{\beta}, \hat{\alpha})=\underset{(\beta, \alpha)}{\operatorname{argmin}} \sum_{i=1}^N(y_i-(\beta_0+\sum_{j=1}^{k}\alpha_{j}\hat{g}_{j}(\bm_i)+\sum_{j=k+1}^{k+p}\beta_j x_{ji}))^2+\lambda \sum_{j=1}^k|\alpha_j|.\end{equation}
$\lambda>0$ is the shrinkage operator, larger values of $\lambda$ decreases the number of models included in the final prediction model.  

When $k$ is large compared to the sample size $N,$ we should use the following loss function  \small
\begin{equation}(\hat{\beta}, \hat{\alpha})=\underset{(\beta, \alpha)}{\operatorname{argmin}} \sum_{i=1}^N(y_i-(\beta_0+\sum_{j=1}^{k}\alpha_{j}\hat{g}_{j}(\bm_i)+\sum_{j=k+1}^{k+p}\beta_j x_{ji}))^2+\lambda_1 \sum_{j=1}^k|\alpha_j|+\lambda_2 \sum_{j=1}^k(\alpha_j)^2\end{equation} \normalsize to allow for more than $N$ non zero coefficients in the final estimation model. $\lambda_1,\lambda_2>0$ are the shrinkage operators.  

In matrix notation, we can rewrite the model in (\ref{eq:additivemodel}) as \[F(X, M; \beta, \alpha)=X\beta+\widehat{G}\alpha\] where $\widehat{G}=(\hat{g}_1,\hat{g}_2,\ldots,\hat{g}_k).$ In our examples, we have used $\widehat{G}\widehat{\alpha}$ as the estimated genotypic values.

It is very important that we note that when using the model in (\ref{eq:additivemodel}) with the hierarchical structure formed by the nested arrangement of genome regions it is almost always better to use all levels at once. 

The authors are also aware that there are other methods which can introduce shrinkage in the parameters like subset selection, partial least squares, principal components regressions, Bayesian lasso, etc... But in essence all these algorithms should give similar results. 

\subsection{Kernels for genomic variables}

In most GWAS studies the focus is on estimating the effects of individual markers and lower level interactions. However, in the genomic era, the number of SNP markers can easily reach millions and the methods used in GWAS for large samples become computationally exhaustive.  The local kernel approach developed in this article remedies this problem by reducing the number of hypothesis by focusing on regions and testing the nested hypothesis in an hierarchy. 

The simplest way we can obtain local kernel matrices is by defining regions in the genome and calculating a separate kernel matrix for each group and region. The regions can be overlapping or discrete. If the some markers are associated with each other in terms of linkage or function it might be useful to combine them together.  The whole genome can be divided physically into chromosomes, chromosome arms or linkage groups. Further divisions could be based on recombination hot-spots, or just merely based on local proximity. We could calculate a separate kernel for introns and exons, non coding, promoter or repressor sequences.  We can also use a grouping of markers based on their effects on low level traits like lipids, metabolites, gene expressions, or based on their allele frequencies. When some markers are missing for some individuals, we can calculate a kernel for the presence and absence states for these markers. When no such guide is present one can use a hierarchical clustering of the variables. It is even possible to incorporate group memberships probabilities for markers so the markers have varying weights in different groups.

\begin{figure}[ht]
\centering
\begin{minipage}[b]{0.45\linewidth}
\includegraphics[width=1\textwidth]{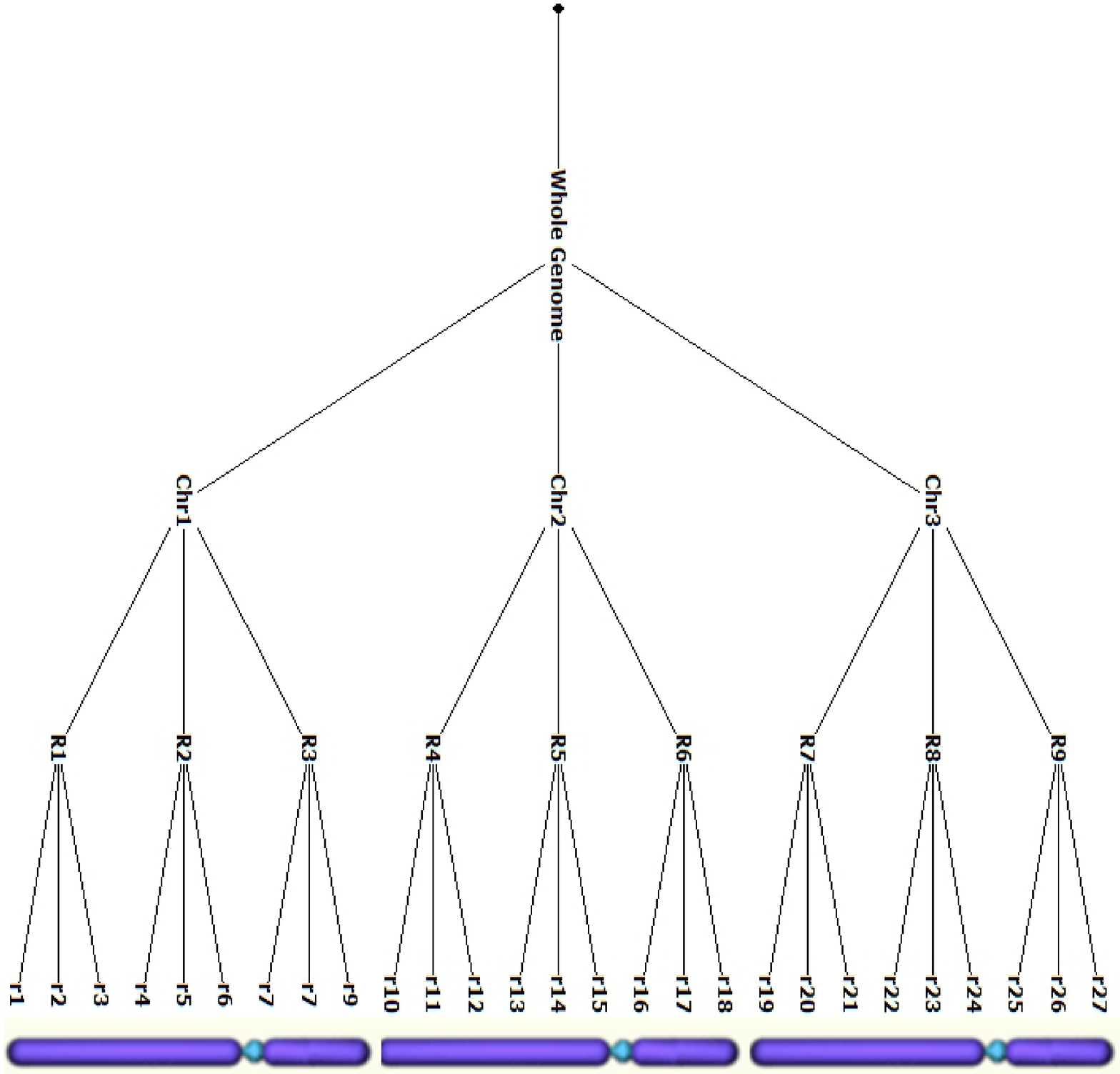}

\caption{An hypothetical hierarchical test set up for an organism with 3 chromosomes. The first test is at the whole genome level. It continues by testing the significance of each chromosome and regions of the chromosome.}
\label{fig:minipage1fighthier}
\end{minipage}
\quad
\begin{minipage}[b]{0.45\linewidth}
\includegraphics[trim=0cm 4cm 0cm 0cm, clip=true,width=1\textwidth]{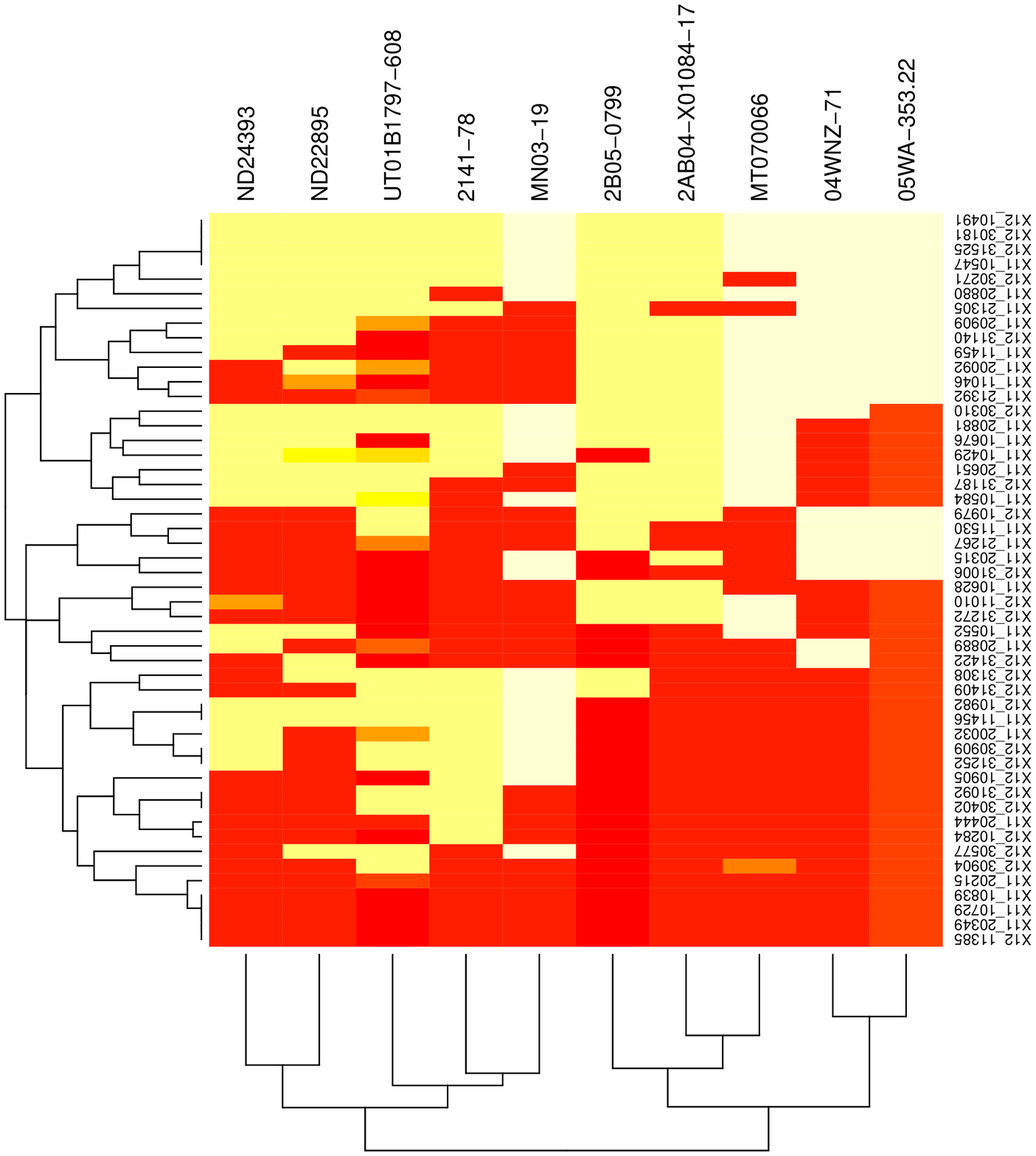}
\caption{When there is no apriori information about the markers, we can use hierarchical variable clustering. This is demonstrated by the clustering on the vertical coordinate of above figure for a random subset of markers and individuals of the FMM data set. }
\label{fig:minipage2clust}
\end{minipage}
\end{figure}

\begin{figure}[htbp]
	\centering	
	\includegraphics[trim=0cm 0cm 2cm 0cm, clip=true,width=1\textwidth]{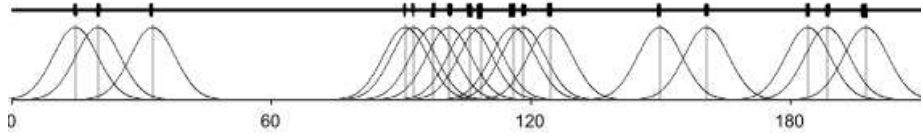}
	\caption{Kernel scanning on a single chromosome with a few markers. At each marker location a relationship matrix is calculated with markers weighted by the kernel weights  obtained from the kernel centered at this marker location. The weights are such that the marker at the center gets the highest weight and the markers get less and less weights as they get away from this center. This is done at each marker location.}
	\label{fig:kernelscanning}
\end{figure}

The second approach which we refer to as kernel scanning requires a linkage map of the markers. This approach is similar to the one in \cite{habier2011extension} where the chromosomes are scanned with windows of 5 consecutive markers. Let $M$ be the $q\times p$ matrix of $p$ markers on $q$ lines, which is partitioned with respect to the chromosomes as $(M_1,M_2,\ldots, M_c)$ where $M_k$ has $p_k$ columns. Let the cumulative distances based on LD between markers in each chromosome be provided in a vector $\bp_k$ for $k=1,2,\ldots,c.$ Based on $\bp_k$ we can to obtain a kernel matrix for markers in each chromosome using a kernel function and by combining these chromosome specific kernel matrices in block diagonal form we obtain a $p \times p$ kernel matrix $S$ for markers. Let the $k$ column of this matrix be represented as $\bs_k.$ A local kernel matrix $K_k$ at position $k$ involves using $diag(\bs_k)^{1/2}M$ in kernel matrix calculations. Kernel scanning approach involves calculation of a kernel matrix for selected marker across the genome at each marker location. By adjusting the kernel width parameter, we are able to determine the smoothness and locality of these kernel matrices. In Figure \ref{fig:kernelscanning} we illustrate kernel scanning on a single chromosome with a few markers.
 
One argument for why we would like to focus on short segments of the genome as distinct structures comes from the ''building blocks'' hypothesis in the evolutionary theory. The schema theorem of Holland \cite{holland1975adaptation} predicts that a complex system which uses evolutionary mechanisms such as fitness, recombination and mutation tend to generate short and well fit structures, these basic structures serve as building blocks. For example, when the alleles associated to an important fitness trait are scattered all around the genome the favorable effects can easily be lost just by independent segregation, therefore inversions that clump these alleles together physically would be strongly selected for.

\subsection{Shrinkage of relationship matrices}

When the number of markers in a region is less than the number of individuals in the training set the kernel matrix for this region becomes singular or ill conditioned. In these cases we can use shrinkage approaches to obtain well conditioned positive definite kernel. The shrinkage estimators of \cite{schafer2005shrinkage} and \cite{ledoit2003improved} that were advocated in \cite{yang2010common} and \cite{endelman2012shrinkage} which involves shrinkage towards the identity matrix are not suitable to use with the SPMM since this involves allocating a fixed proportion of the error variance to the variance of the random effects. We instead propose and use shrinkage estimators which aim to introduce sparsity in the off diagonal elements of the kernel matrix. Many algorithms have been devised for learning sparse covariance matrices in the recent years. A penalized maximum likelihood estimation was developed in \cite{friedman2008sparse}, a penalized regression method was used in \cite{meinshausen2006high}. These and some other sparse covariance estimation techniques are implemented in an R package ''huge'' (\cite{zhao2012huge}). 

\begin{figure}[htbp]
	\centering	
	\includegraphics[trim=0cm 1.5cm 0cm 2cm, clip=true, width=.5\textwidth]{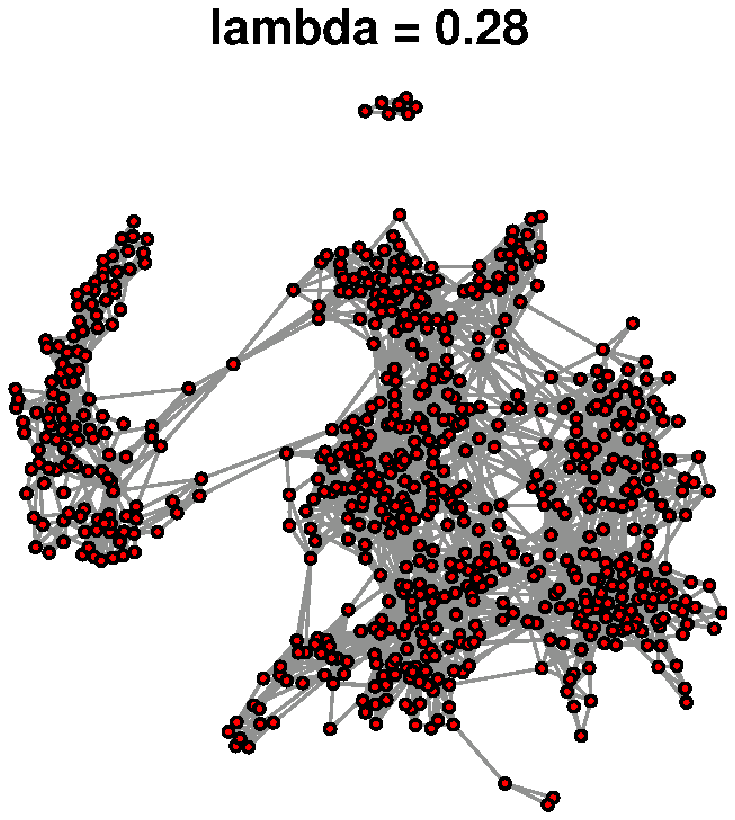}
	\caption{Barley CAP. The different plants are represented by the dots and the nonzero relationship coefficients are represented by the lines between them. The graphical representation of the shrunk relationship matrix gives us an idea about the structure of the Barley CAP population.}
	\label{fig:sparsecov}
\end{figure}

In addition to possible decrease of computational burden and increase in accuracy by use of sparse matrix methods in mixed model parameter estimation, we can produce graphical representations of the kernel matrices. For a normally distributed random vector, the independence between two components is implied by zero covariance between the components, more interestingly, the conditional independence between two components is implied by the zero components in the inverse covariance matrix. In Figure \ref{fig:sparsecov}, we display the graphical representation of the sparse realized relationship matrix.

\section{Illustrations}

In this section, we will compare the methods introduced in this article to some of the existing ones. Our first example uses simulated markers and phenotypes and so the truth is known. The remaining examples are reserved for barley data sets we have downloaded from the Triticaea Toolbox web portal at  \url{https://triticaeatoolbox.org}.


\begin{ex}
The data in this example was generated by a whole genome simulator ''hypred'' \cite{liang2007genome} which is an R package that simulates high density genomic data.  Markers for each of the 7 chromosomes of length 1M are simulated for individuals which were produced randomly mating two founder lines for 20 generations. The total number of markers was 3000. On each chromosome 20 QTL positions and additive effects were randomly generated. Residual variance was set to adjust the heritability of the trait to 0.75. All marker effects are additive. The number of individuals is set to 1000. 750 of these individuals were randomly chosen to the training set and remaining to the test data set. In Figure \ref{testtraingaussian1}, we compare the accuracies in the test data using the correlation scores between the observed and predicted trait values for the linear and Gaussian kernel models with the multiple kernel models that divide the genome into pieces differing number of regions. Number of regions were set to $2^2=4$, $3^2=9$,...,$15^2=225$ (two levels of hierarchy).  The experiment was repeated 30 times.

\begin{figure}[htbp]
	\centering
		\includegraphics[width=.7\textwidth, angle=270]{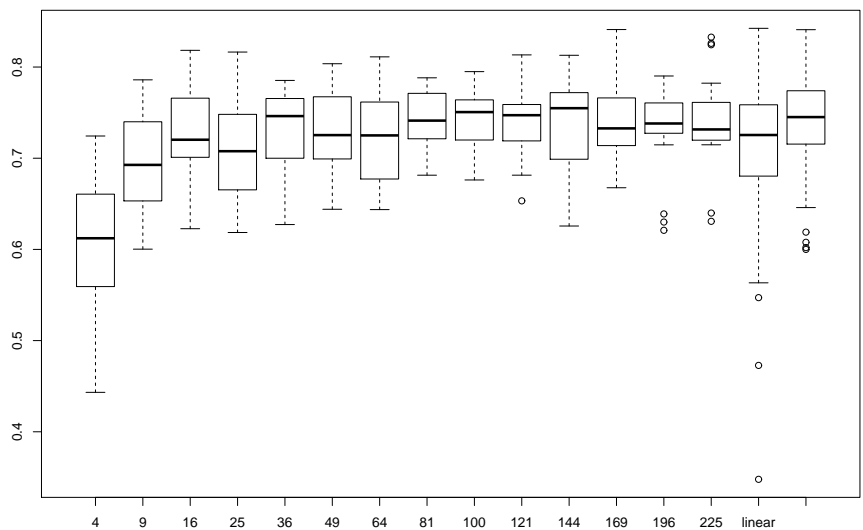}
		\caption{Cross validated accuracies measured in terms of correlation scores for simulated data described in Example 1.}
		\label{testtraingaussian1}
\end{figure}
In Figure \ref{fig:HKVplot}, we displayed the local heritability values from the kernel scanning approach. Finally, in Figure \ref{fig:Regionsandfit}, we display the results from a hierarchical testing procedure. 

\begin{figure}[htbp]
	\centering
		\includegraphics[width=1.00\textwidth]{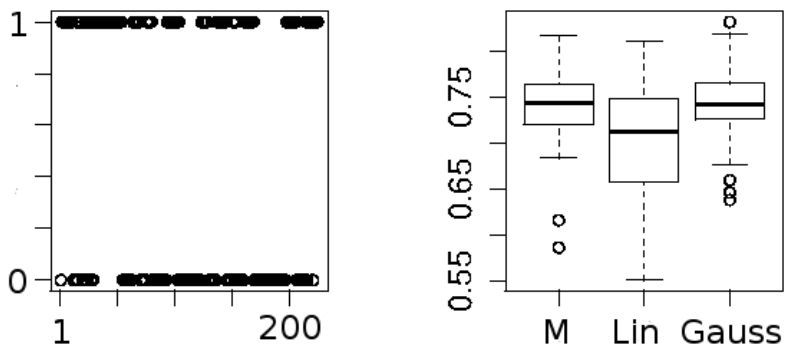}
	\caption{The figure on the left illustrates the sparsity pattern obtained using the hierarchical testing procedure for one instance of the experiment  described in Example 1 We have used the hierarchical testing procedure of Meiwaussen to identify the relevant regions of the genome for the data in Example 1. Only the regions with significant effects were used to build the multiple kernel model (regions with 0 value are not included in the final model). This model (M) is compared with the SPMM's with linear and Gaussian kernels (Lin and Gauss) with the boxplots on the right. All the models have approximately the same prediction ability, but the multiple kernel model is definitely more parsimonious (parts of genome included in the final prediction model are indicated by ones on the left graph).}
	\label{fig:200qtl}
\end{figure}

\begin{figure}[htbp]
	\centering
\includegraphics[angle=0,width=1\textwidth]{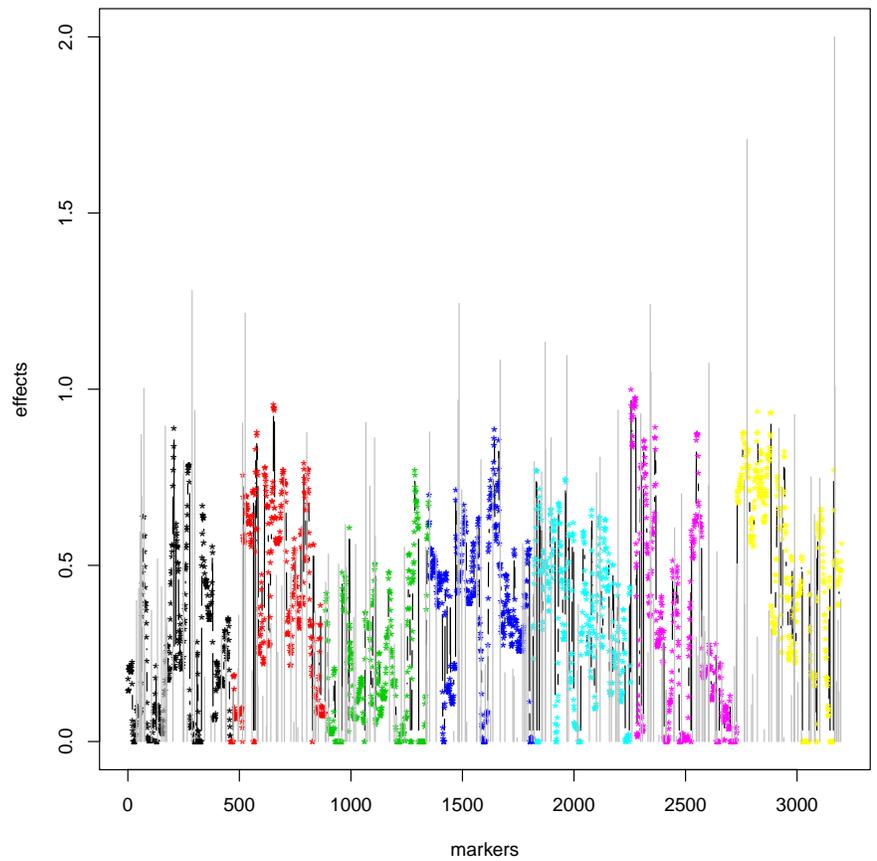}
	\caption{The local heritability values from the kernel scanning approach. The true QTL and effect sizes are superimposed on the estimated regional effects (Horizontal bars for true effects, colored points for estimated effects in chromosomes).}
	\label{fig:HKVplot}
\end{figure}

\end{ex}

\begin{ex} In an experiment carried out by USDA-ARS  during the years 2006-2007, the alpha-tocopherol levels for 1723 barley lines were recorded in total of 4 environments (2 years and 3 locations). Along with the phenotypic information 2114 markers on 7 chromosomes (unmapped markers were assigned to an arbitrary 8th chromosome) were available for the analysis. The whole genome was divided in a similar fashion as displayed in Figure \ref{fig:minipage1fighthier}.  We have sampled 500 lines for training the models and we have used the rest of the lines to  evaluate the fit of our models.  In particular, we have calculated the correlation between the phenotypic values in the test data and the corresponding estimated genotypic values from our models. This was repeated 30 times and the models are compared in Figure \ref{fig:minipage1}. The lasso importance scores obtained for the genome regions from models at 4 hierarchical levels is given in Figure \ref{fig:minipage2} (the displayed weights are averaged over the 30 replications of the experiment).

\begin{figure}[ht]
\centering
\begin{minipage}[b]{0.45\linewidth}
\includegraphics[angle=270,width=1\textwidth]{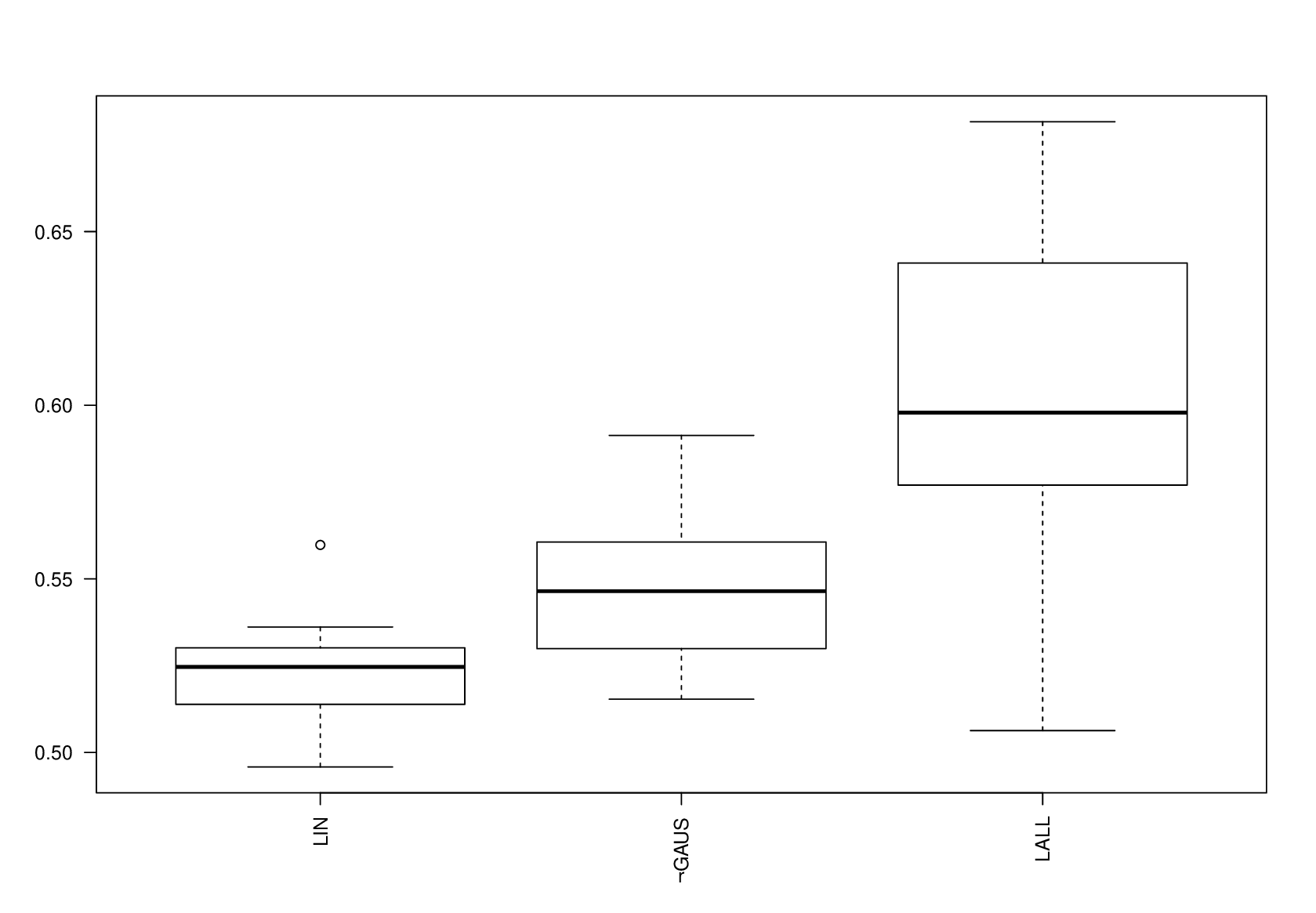}

\caption{Three models (1-SPMM with linear kernel, 2-SPMM with Gaussian kernel, 3- lasso model in \ref{eq:additivemodel} using estimated random effects from all levels) are compared based on the correlation between the phenotypic values in the test data and the corresponding estimated genotypic values from our models.}
\label{fig:minipage1}
\end{minipage}
\quad
\begin{minipage}[b]{0.45\linewidth}
\includegraphics[angle=270,width=1\textwidth]{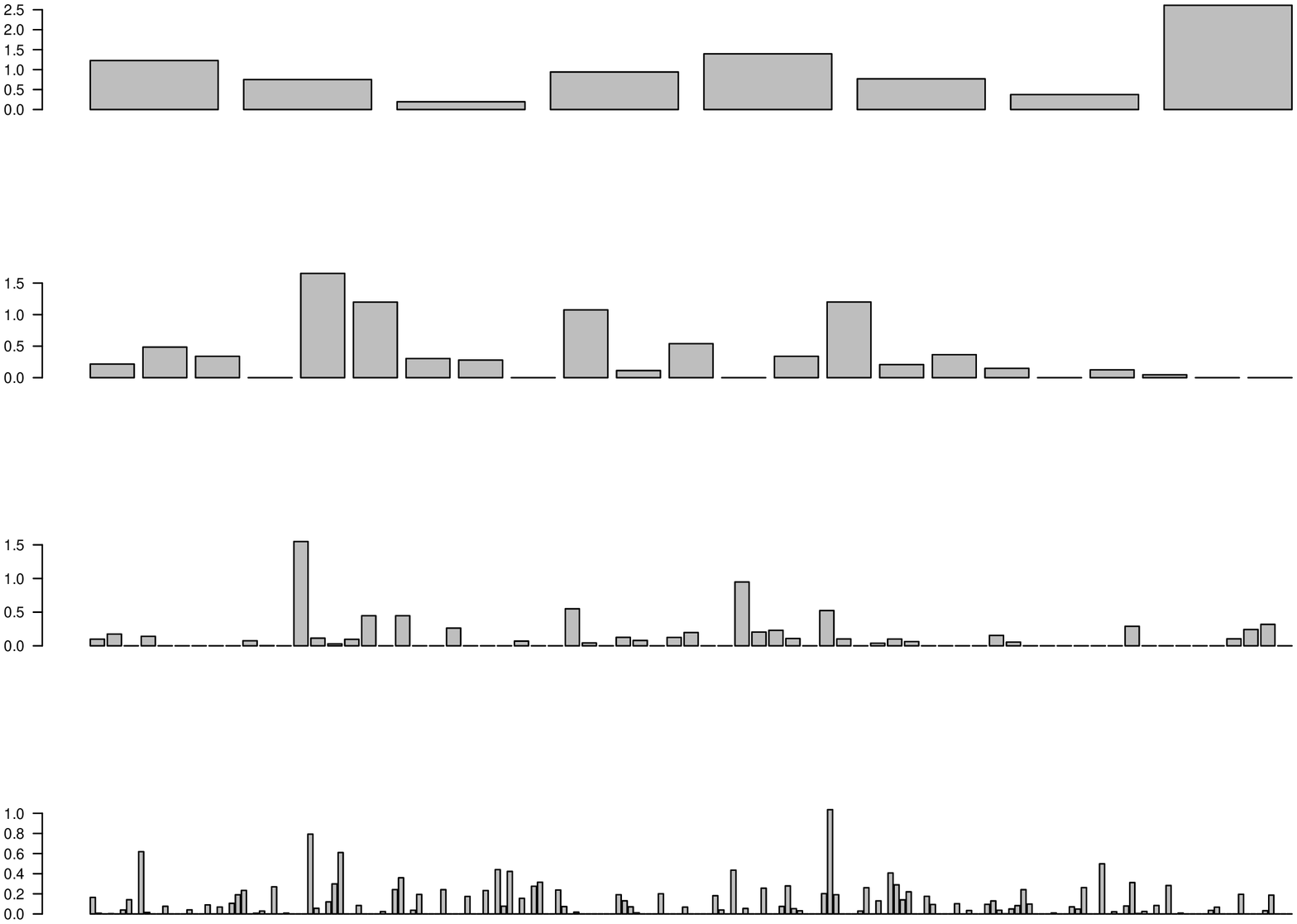}
\caption{The lasso importance scores obtained for the genome regions from models at 4 hierarchical levels (the displayed weights are averaged over the 30 replications of the experiment). }
\label{fig:minipage2}
\end{minipage}
\end{figure}

\end{ex}

While we may have confidence that GS can accelerate short-term gain, no such confidence is justified for long-term gain.  Beyond the first cycles of selection, mechanisms the effects of which are difficult to predict analytically begin to operate. In \cite{goddard2009genomic} and \cite{jannink2010dynamics} a weighted GS model was used so that markers for which the favorable allele had a low frequency should be weighted more heavily to avoid losing such alleles. In this article we recommend using the similar approach which aims to conserve rare but favorable alleles for balancing short-term and long-term gains from GS. Let $\hat{g}_{ji}$ for regions $j=1,2,\ldots, k$ and individuals $i=1,2,\ldots, N$ be the EBLUPs of random effect components that correspond to the $k$ local kernels and let $\hat{p}_{ji}$ denote the estimated density value of the alleles in region $j$ for individual $i.$  A selection criterion in the spirit of \cite{jannink2010dynamics} for the $i$th individual is then \begin{equation}\hat{c}_i=\sum_{j=1}^{k}\frac{\hat{g}_{ji}}{1+\sqrt{\hat{p}_{ji}}}.\end{equation} However, this criterion also down weights favorable but common alleles. 

A better approach is using a weighting scheme that can up weight both rare and common favorable alleles based on breeders preferences. Let $\hat{g}_{j(n)}$ be the maximum EBLUP value among the individuals for region $j$ and let $\eta_j$ be the weight of the kernel $j.$ The selection criterion \begin{equation}\label{selind}\hat{c}_i=\sum_{j=1}^{k}\frac{\hat{\eta}_j}{2\pi h_1 sd(\hat{g}_{j})h_2sd(\hat{p}_j)}\exp{\{-\frac{1}{2}\left[\frac{(\hat{g}_{ji}-\hat{g}_{j(n)})^2}{h_1^2 var(\hat{g}_{j})}+\frac{\hat{p}^2_{ji}}{h_2^2var(\hat{p}_j)}\right]\} }\end{equation}  where the constants $h_1$ and $h_2$ are selected by the breeder based on preferences given to short term and long term gains correspondingly.  

\begin{figure}[htbp]
	\centering
		\includegraphics[width=1\textwidth]{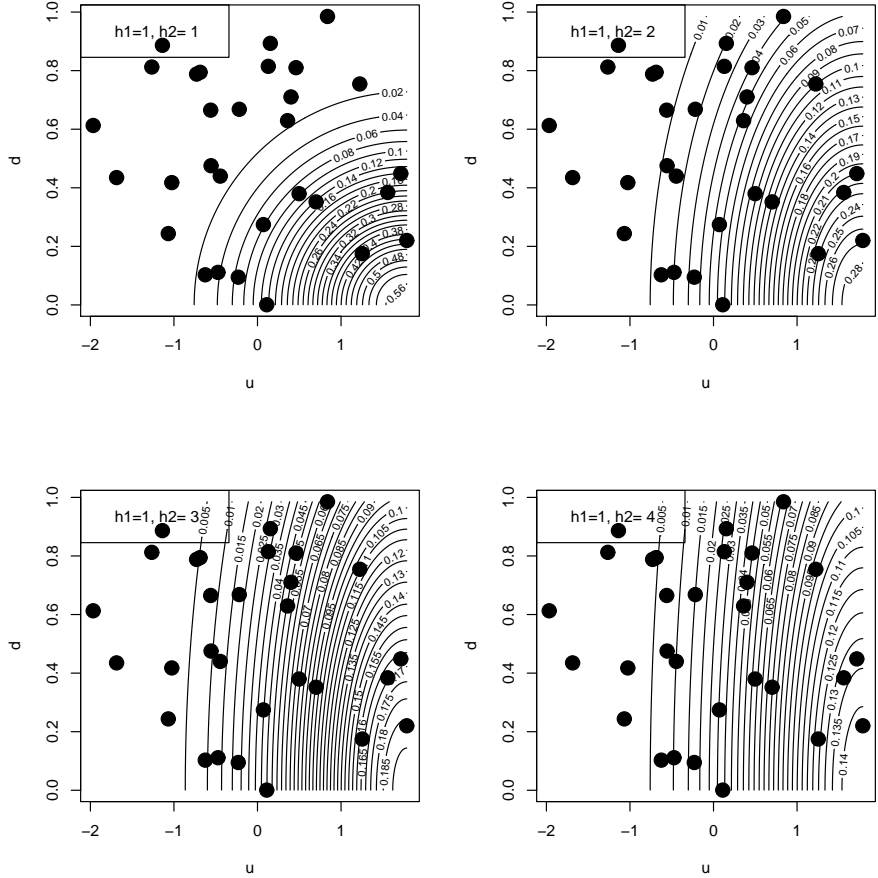}
	\caption{Levels of the contours represent the value of the selection index in formula (\ref{selind}). As $h_2$ increases the weight put on the density decreases and the selection index gives the same ordering as the ordinary GEBVs.}
	\label{fig:weightedblupscontours}
\end{figure}

More importantly the estimates of local heritabilities and the estimates for the local effects for the lines in the breeding population can be used to approximate the distribution of the phenotype for the individual crosses. This information is as important to the breeder as a good prediction model since it gives a guide for action. 

Let $\hat{\bg}_1=(\hat{g}_{j1})_{j=1}^{k}$ and $\hat{\bg}_2=(\hat{g}_{j2})_{j=1}^{k}$ be the EBLUP values of two individuals for the $k$ regions of the genome. A typical cross between two double haploid individuals will have a breeding value in the form \begin{equation}[0,1,0,0,\ldots,0,1]'\hat{\bg}_1+[1,0,1,1,\ldots,1,0]'\hat{\bg}_2\end{equation} since each region is inherited from either of the parents. If we can assume the assortment is independent between each region and with equal chance an approximation of the distribution of the phenotype for the off springs of these plants can be obtained by Markov chain Monte Carlo techniques. The law of independent assortment always holds true for genes that are located on different chromosomes, but for genes that are on the same chromosome, it does not always hold true.

\section{Conclusions}

The approaches introduced allows us to use the input variables in naturally occurring blocks. In the context of the SPMM in (1) there are very fast algorithms that can take advantage of this dimension reduction. For the linear kernel function, the order of calculations to solve a SPMM with one kernel matrix is proportional to $min(n,m)$ where $m$ here is the number of features in that kernel. No matter what the input dimension is SPMM parameter estimation involves matrices of order $n.$   Therefore, the multiple kernel approach overcomes the memory problems that we might incur when the number of markers is very large. 

The local kernels use information collected over a region in the genome and, because of linkage, will not be effected by a few missing or erroneous data points, so this approach is also robust to missing data and outliers.  

For QTL identification, we have recommended a nested sequential approach that levels of views of the genome which might lead to faster exploration of the whole genome for quantitative trait loci. Recently popular deep learning algorithms try to learn levels of attributes and hypothesis \cite{bengio2009learning}. Also active learning algorithms search for the parts of the data to obtain information in a stepwise fashion \cite{settles.tr09}. ''Testing along a tree of hypotheses'' approach to association studies is related to these main stream methods of machine learning. 

Sexual gene transfer methods have been used successfully in plant breeding for thousands of years.  More recently, nonsexual  methods have also been incorporated. The multiple kernel mixed model approach allow us to evaluate the utility of genome regions. This, in turn, allow us to build models with good prediction accuracy and, more importantly, aides us in our action: Which plants,  chromosomes or genome regions should be kept in our breeding program? Which crosses are most useful? Which genome regions should be transfered between individuals? Success with genomic selection partially depends on good prediction models and partially on utilization of this kind of information and new emerging technologies.

Although we have focused our attention to classical breeding with crosses among selected parents, a short cut to breeding better plants and animals is possible by isolation and fusion of individual chromosomes or genomic regions.  This type of breeding, which we call chromosomal breeding, involves to breeding better chromosomes and combining them. The authors believe that the plants obtained by passing chromosomes within species or families involve minimal genetic manipulation.

\section*{Acknowledgments}
This research was supported by the USDA-NIFA-AFRI Triticeae Coordinated Agricultural Project, award number 2011-68002-30029.

\bibliographystyle{plain}

\bibliography{kernelbib.bib}

\end{document}